\documentstyle[emulateapj]{article}
 
\slugcomment{(Submitted to ApJ)}
 
\lefthead{J.Koda, Y.Sofue and K.Wada}
\righthead{Origin of the Tully-Fisher Relation}

\newcommand{\Msun}{M_{\odot}}
\newcommand{\kpc}{{\rm \, kpc}}
\newcommand{\mpc}{{\rm \, Mpc}}
\newcommand{\magni}{{\rm \, mag}}
\newcommand{\kmps}{{\rm \, km \, s^{-1}}}

\begin{document}
\title{On the origin of the Tully-Fisher relation}

\author{Jin Koda$^1$, Yoshiaki Sofue$^1$ and Keiichi Wada$^2$}
\affil{1.Institute of Astronomy, University of Tokyo, Mitaka, Tokyo 181-8588, Japan\\
2.National Astronomical Observatory, Mitaka, Tokyo 181-8588, Japan
}

\begin{abstract}
We discuss the origin of the Tully-Fisher (TF) relation using the $N$-body/SPH
method, which includes cooling, star formation and stellar feedback
of energy, mass and metals. We consider initially rotating
overdense spheres, and trace formation processes of disk galaxies from
$z=25$ to $z=0$ in the Cold Dark Matter (CDM) cosmology.
To clarify the origin of the TF relation, we simulate formation
of 14 galaxies with different masses and spin parameters, and compute
observable values, such as the total magnitude and the line-width. We find
that the simulated galaxies reproduce the slope and scatter of the TF
relation: the slope is originated in the difference of total galactic masses,
and the scatter is produced by the difference of initial spin parameters.
As well as the TF relation, observed features of spiral galaxies, such as
the exponential light-profile and the flat rotation curve, are reproduced
in our simulations, which were assumed {\it a priori}
in past semi-analytical approaches.
\end{abstract}


\section{INTRODUCTION}\label{sec:intro}
The Tully-Fisher relation (hereafter TF: Tully \& Fisher 1977) is
one of the most basic relations in spiral galaxies, and would be a clue
to understand the origin of disk galaxies. The TF relation has been roughly
considered as a product of the virial theorem and a nearly constant
mass-to-light ratio since its discovery. Recently, the origin of the TF
relation is discussed by, e.g., Silk (1997) and Mo, Mao \& White (1998).

In their semi-analytical approach, Mo, Mao \& White (1998) succeeded in
reproducing the TF relation, assuming a constant mass-to-light ratio
and empirical profiles of disks and haloes.
Heavens \& Jimenez (1999) took the same approach but including an empirical
star formation model, and reproduced the TF relation in four passbands
simultaneously. In these semi-analytical approaches, however, observed
features of galaxies, such as the exponential profile and flat rotation
curve, were not constructed as the results of simulations, but assumed
{\it a priori}. Steinmetz \& Navarro (1999) performed direct simulations of
galaxy formation within the cosmological context, and also succeeded in
getting the slope and scatter of the TF relation. They treated the volume
much larger than the scale of galaxies, and considered environmental effects
(e.g., tidal field and infall/outflow of mass). However, due to
the complicated behaviors of these effects, it still remains unknown
what physics produces the TF relation.

Based on these arguments, we simulate formation and evolution of galaxies
with the $N$-body/SPH method including cooling, star formation
and their feedback of energy, mass and metals to the ISM. Using the similar
method, Katz (1992) and Steinmetz \& M\"uller (1994, 1995) succeeded
in constructing internal structures as observed in spiral galaxies,
e.g., the exponential density profile, flat rotation curve,
and distributions of stellar age and metallicity in the bulge, disk and halo.
In contrast to their simulations of single galaxy formation, we consider
formation of many galaxies with different masses and spin parameters.
In order to investigate what physics produces the TF relation, we concentrate
on formation of isolated galaxies with different initial conditions.
We note that as well as the TF relation, observed features of spiral galaxies,
such as the exponential light-profiles and flat rotation curve, are reproduced
in our simulations, which were assumed {\it a priori} in past
semi-analytical approaches.

\section{METHODS}
\subsection{Numerical Methods}
We use a GRAPE-SPH code, a hybrid scheme of the smoothed particle
hydrodynamics (SPH) and the $N$-body integration hardware GRAPE-3
(Sugimoto et al. 1990), the code which was firstly discussed by
Steinmetz (1996) and developed further by us (Koda, Wada \& Sofue 1999).
The code is implemented to treat three-component systems: the gas, stars and
dark mater, and calculate gravitational and hydrodynamical forces, which
are symmetrized to satisfy the Newton's third law. 
We consider the radiative and inverse Compton cooling of the gas
which is optically thin and in collisional ionization equilibrium of H
and He ($X_H=0.76$, $X_{He}=0.24$; see Katz et al. 1996).
We also take phenomenological models of star formation and their feedback
to mimic real galaxy formation.

The star formation model adopted is basically the same as those discussed in
Katz (1992), Navarro \& White (1993) and Steinmetz \& M\"uller (1994, 1995).
Stars are formed in regions which are locally contracting and Jeans-unstable
at a rate given by $\dot{\rho_\star} = c_\star \rho_{\rm gas}/\max
(\tau_{\rm dyn}, \tau_{\rm cool})$. Here, $\rho_\star$, $\rho_{\rm gas}$,
$\tau_{\rm dyn}$ and $\tau_{\rm cool}$ are the densities of stars and the gas,
local dynamical and cooling timescales, respectively. In most cases,
$\tau_{\rm dyn}$ is larger than $\tau_{\rm cool}$, and the star formation
timescale is typically $\sim 20 \tau_{\rm dyn}$ under the adopted parameter
$c_\star=0.05$. When one third of the mass in a gas particle is transformed
into stars, we create a new collisionless {\it star particle}, which inherits
the position and velocity of its parent gas particle.

We take a simple model of feedback. Massive stars with $\geq 8 \Msun$ are
assumed to release energy, mass and metals into the surrounding gas
at a constant rate through stellar wind (SW) and type II supernovae (SNII).
All of the massive stars then become white dwarfs, and $15 \%$ of them
results in type Ia supernovae (SNIa, Tsujimoto et al. 1995).
In Table \ref{tab:eject}, we list total released energy, mass and metal
per star, and the periods of feedback (see Nomoto et al. 1997a \& 1997b;
Yoshii et al. 1996). We can estimate the number of stars with $\geq 8 \Msun$,
in a {\it star particle}, with the initial mass function (IMF) of Salpeter
(1955). Released energy is provided into the surrounding gas
as thermal energy. Since stars are formed and release energy at dense gas
regions, where feedback energy is soon radiated away, feedback does not so
much affect the simulations in our model.

\subsection{Initial Conditions}
We simulate formation and evolution of galaxies from $z=25$ to $0$ in the
CDM cosmology ($\Omega_0=1$, $h = 0.5$). Our initial conditions are similar
to those of Katz (1992) and Steinmetz \& M\"uller (1994, 1995), but we consider
14 isolated spheres, on which small scale CDM fluctuations are superimposed
with the Zel'dovich approximation (Zel'dovich 1970). The densities of
the spheres are enhanced above the background field by $\delta \rho/\rho=0.25$.
We normalize the CDM spectrum so that the rms
fluctuation in a sphere of radius $8 h^{-1} \mpc$ becomes equal to
$\sigma_8=0.63$ at $z=0$. The spheres are rigidly rotating, and following
the reduced Hubble expansion (see Steinmetz \& M\"uller 1995).
Two free parameters, mass and spin parameter,
are listed in Table \ref{catalog}.

The gas and dark matter are represented by the same number of particles,
and their mass ratio is set to $1/9$. The initial temperature
of the gas is set to 70 K of the cosmic microwave background at $z=25$.
The mass of a gas particle varies between $2.4 \times 10^{6}$ and $1.9 \times
10^{7} \Msun$ according to the system mass considered (Table \ref{catalog}).
The mass of a dark matter particle varies between $2.1 \times 10^{7}$ and
$1.7 \times 10^8 \Msun$. Low resolution may cause an artificial heating due
to two-body relaxation between the SPH and dark matter particles,
however this range of particle mass is small enough to exclude
the artificial heating effect (Steinmetz \& White 1997).
The gravitational softenings are taken to be $1.5 \kpc$ for gas and star
particles, and $3 \kpc$ for dark matter. The total number of particles
typically becomes $\sim 4 \times 10^{4}$ at the end of the simulations ($z=0$).

\subsection{Data Reduction}
In order to compare the properties of observed and simulated galaxies,
we compute the observables such as luminosity and the line-width for each
``spiral galaxies'' at $z=0$. Stellar luminosity is computed
with the simple stellar population (SSP) synthesis models of
Kodama \& Arimoto (1997). The models provide integrated spectra along
stellar isochrones corresponding to appropriate age and metallicity.
We take the Salpeter's IMF. Total magnitude $M_{I}$ of simulated
galaxies is computed to sum up star particles using appropriate SSP
tables, which are selected in accordance with age and metallicity of
the star particles. The line-width $W_{20}$ is derived
similarly to the observable as constructing a line-profile of gas
weighted by mass, and measuring the width at $20 \%$ level of a peak
flux. The catalog of all the simulated galaxies is presented in
Table \ref{catalog}.

\section{RESULTS}
\subsection{Evolution and Structures of Individual Galaxies}
\begin{figure}
\plotone{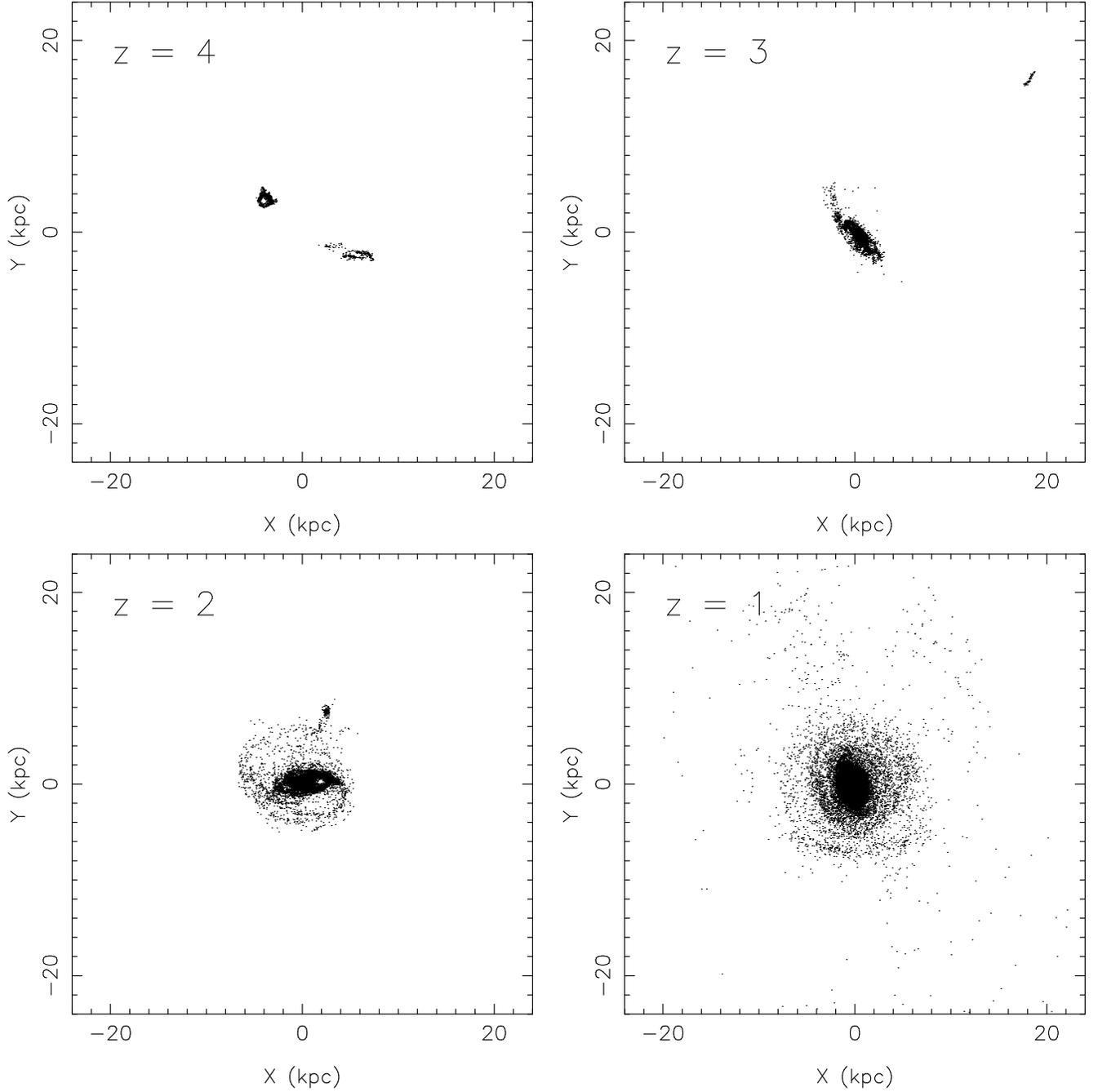}
\figcaption{Snapshots of star particles at various redshifts in the 
case for $M=4 \times 10^{11} \Msun$ and $\lambda =0.06$.
The origin of each figure is shifted to make the object nearly at the center.
The view angles gives the face-on projection at $z=0$.
Note that these plots do not actually show the mass density because
each star particle does not necessary have the same mass.
\label{fig:snap_evol}}
\end{figure}

Figure \ref{fig:snap_evol} shows snapshots of star particles at four
redshifts for the case of $M=4 \times 10^{11} \Msun$ and $\lambda=0.06$.
These view angles provide a face-on projection at $z=0$.
Two clumps at $z=4$ are merging and form a bulge-like system between
$z=4$ and $3$. After the bulge formation, the surrounding gas gradually
cools and falls to form a gaseous disk, and then a stellar disk is gradually
formed. The panel at $z=2$ shows that a disk-like structure surrounds
the central bulge. Since the gas density of the inner disk
is higher than that of the outer part, the star formation timescale is shorter
in the inner part than in the outer part. The stellar disk is gradually
formed from the inside to outside in the gas disk. The panel at $z=1$ shows
a larger disk than that of $z=2$.

\begin{figure}
{\epsscale{0.7}
\plotone{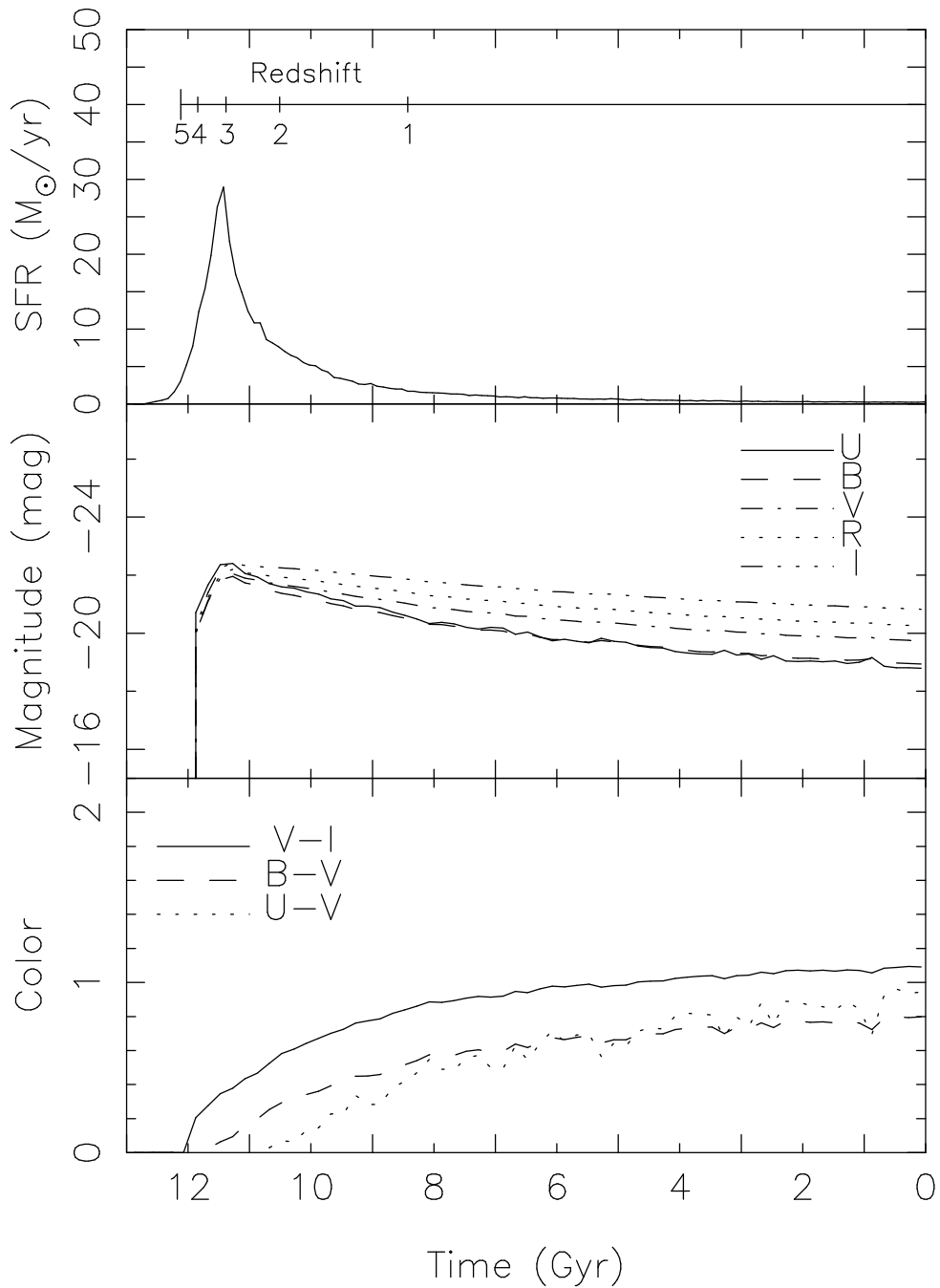}
\figcaption{
Evolution of total star formation rate, total magnitude and color in the case
for $M=4 \times 10^{11} \Msun$ and $\lambda =0.06$.
{\it Upper panel:} star formation rate vs time.
{\it Middle panel:} U, B, V, R and I-band total magnitude in a rest frame
vs time.
{\it Lower panel:} V-I, B-U and U-V color in a rest frame vs time.
The scale in redshift is also illustrated in the upper panel.
\label{fig:parm_time}}
}
\end{figure}

In Figure \ref{fig:parm_time} we plot the total star formation rate,
total magnitude and color in a rest frame against the look-back time
(and also the redshift). The star formation rate peaks at $z \sim 3$
at a rate of $29 \Msun {\rm yr}^{-1}$, and then declines and reaches
an almost constant value of $1 \Msun {\rm yr}^{-1}$ at $z \sim 1$,
when the stellar mass has already been about $80\%$ of the total disk mass.
At $z=0$, about $10\%$ of the total baryonic mass remains in the gas and
the rest is already in stars, which is consistent with the observations of
spiral galaxies. Total magnitude also peaks near the time of the maximum star
formation. Then the magnitudes in all passbands are monotonically
declining until $z=0$ because the mean age of the stellar component
becomes older and massive stars gradually die. The amount of the decline
between $z=3$ to the present is $1.6 \magni$ in $I$-band.

\begin{figure}
{\epsscale{0.7}
\plotone{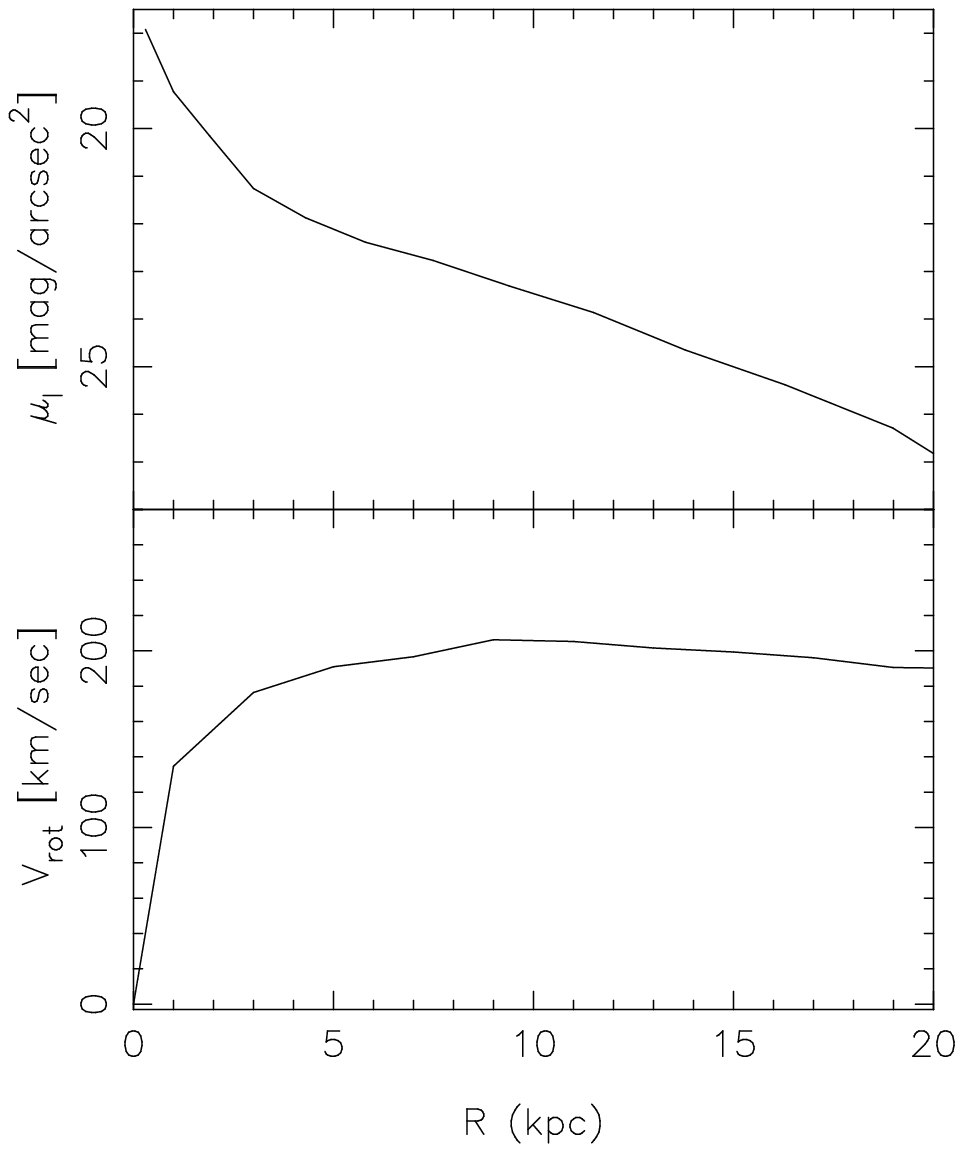}
\figcaption{Internal structures of final galaxies at $z=0$ in the case
for $M=4 \times 10^{11} \Msun$ and $\lambda =0.06$.
{\it Upper panel:} the $I$-band surface brightness $\mu_I$ profile.
{\it Lower panel:} the rotation curve.
All the simulated galaxies finally show the exponential light-profiles and
flat rotation curves, similar to observed spiral galaxies.
\label{fig:profile}}
}
\end{figure}

We show the final snapshots of star particles in Figure \ref{fig:snap_z=0}.
The $I$-band luminosity profile and rotation-velocity profile
of the final galaxy are shown in Figure \ref{fig:profile}.
Observed internal structures, such as the exponential light-profile
and flat rotation curve, are well reproduced as the results of
the simulation. We confirm that all the 14 simulated ``spiral galaxies''
attain such properties.

About $30\%$ of total gas angular momentum transfers to that of dark matter
in the period $z=5-2$, when condensed gas core with dark halos merge into
a larger object (Katz \& Gunn 1991; Navarro, Frenk \& White 1995).
After the period, the baryon disk scarcely looses the angular momentum,
because of no infall of lumpy clumps in our simulations.
The specific angular momenta of the baryon disks lie in the range of
those in observed spiral galaxies at $z=0$ (Fall 1983; Contardo et al 1998).
In the following discussion, we investigate a statistical property of
14 simulated galaxies, all of which have internal structures similar
to observed spiral galaxies as discussed above.

\subsection{The Tully-Fisher Relation}\label{sec:TF}
\begin{figure}
\plotone{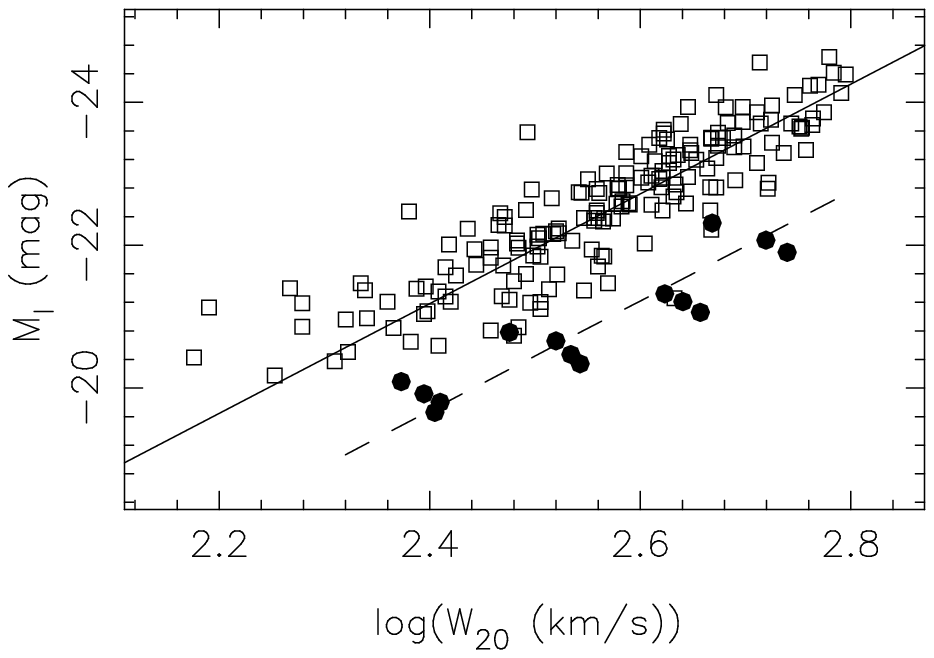}
\figcaption{The TF relation of observed (open squares) and
simulated galaxies (filled circles). The slopes of the solid and dashed lines
are derived by fitting to observed data, and the zero-point are fitted by
eye. Simulated galaxies well reproduce the slope and scatter of the TF
relation except for their $1.5 \magni$ fainter zero point.
\label{fig:tf_comp}}
\end{figure}

In Figure \ref{fig:tf_comp}, we compare the observed (open squares)
and simulated (closed circles) TF plot in $I$-band.
We use the observed data presented by Han (1992) and translate them to
absolute values with distances assuming $h=0.5$.
The slopes of the solid and dashed lines are derived by fitting to observed
data (Giovanelli et al. 1997a), and the zero points are fitted by eye.
In Figure \ref{fig:tf_comp}, we notice that the simulated TF relation shows
the following three points:
(i) The slope of TF is well reproduced.
(ii) The scatter of TF is also similar to the observations.
(iii) The zero point is systematically fainter.
These points are consistent with the previous results which included
environmental effects, e.g., tidal field and infall/outflow of mass
(Steinmetz \& Navarro 1999). Here, we artificially control initial conditions
of isolated spheres, and examine how final structures of ``spiral'' galaxies
depend on the initial conditions. Based on these simulations, we discuss
the TF relation in detail on three points: the zero point, slope and scatter.

{\it The zero point}:
Our simulated galaxies are $1.5 \magni$ fainter than observed ones.
We could find a possible solution of this discrepancy by changing
the adopted cosmological model: Changing $h$ ($\Omega_0$) would
cause a vertical (horizontal) shift of the plotted points
in Figure \ref{fig:tf_comp}. If larger $h$ ($>0.5$) is taken, distance
estimations for the observed galaxies become smaller, and open squares
(observed galaxies) vertically shift downward in Figure \ref{fig:tf_comp}.
Larger $h$ also provides younger age, and therefore, brighter luminosity
for galaxies, because galaxies loose luminosity with their age.
It results in an upward shift of the simulated galaxies.
If smaller $\Omega_0$ ($<1$) is taken, the baryon fraction
$\Omega_{\rm b}/\Omega_0$ becomes larger because $\Omega_{\rm b}$ should be
fixed by the Big Bang Nucleosynthesis. Since $I$-band luminosity is almost
determined by the total baryonic mass, it results in a smaller mass-to-light
ratio $M/L$, which shifts the filled circles (simulated galaxies) leftward in
Figure \ref{fig:tf_comp}. If we take a cosmological model with larger $h>0.5$
and/or smaller $\Omega_0<1$, which is suggested by recent observations
(Giovanelli et al. 1997b, Perlmutter et al. 1998), the simulated TF relation
would become consistent of the observed one. Note that the zero point
also depends on the redshift when the whole mass is merging into
a galaxy in hierarchical cosmology (Mo, Mao \& White 1998).
The $\Omega_0<1$ universe will shift the zero point brighter (upward),
and also alleviate the discrepancy.

\begin{figure}
{\epsscale{1.0}
\plotone{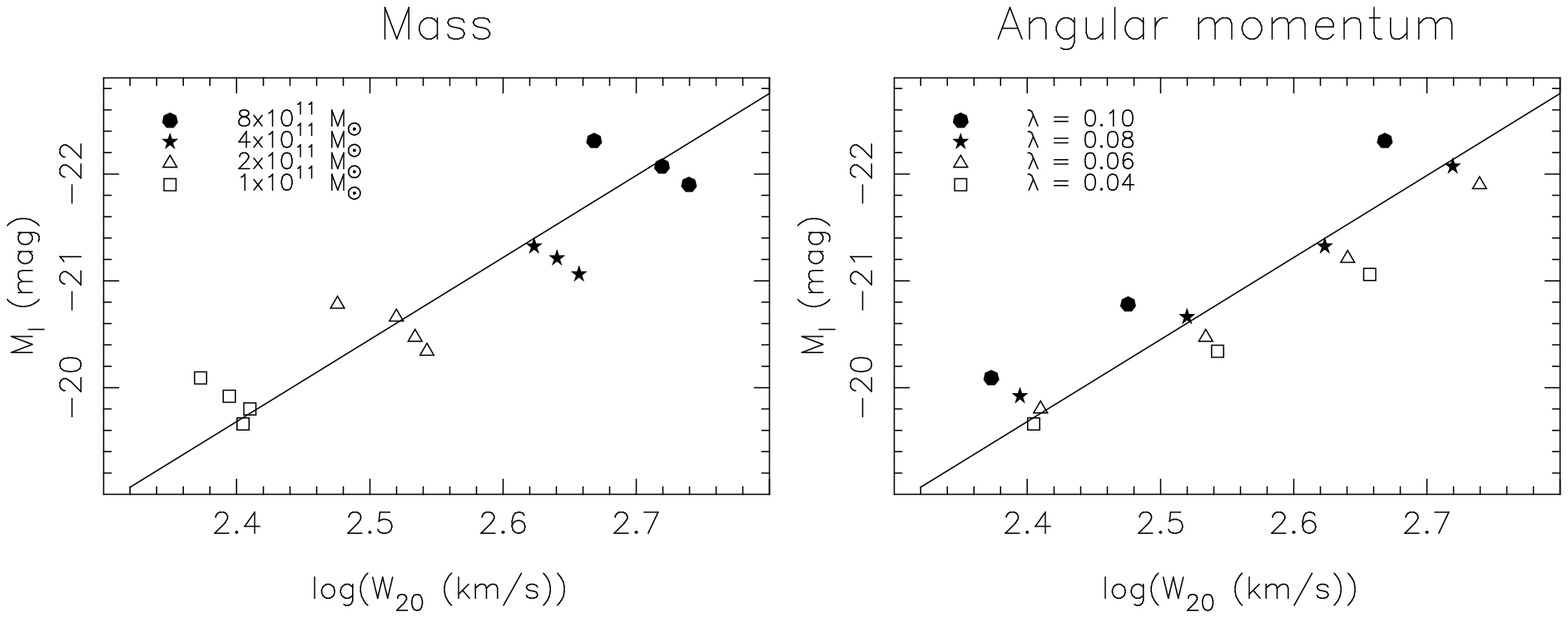}
\figcaption{
The TF relations of simulated galaxies. The slopes of the solid lines are
derived by fitting to observed data, and the zero-point is fitted by eye. 
{\it Left panel:} mass dependence of galaxy distribution in the TF plot.
Different masses are explicitly shown with different symbols.
The TF relation is extended to the direction which traces different masses.
{\it Right panel:} angular momentum dependence of galaxy distribution
in the TF plot.
Different spin parameters $\lambda$ are shown with different symbols.
Galaxies with different $\lambda$ distribute nearly perpendicular to
the TF relation, and result in the scatter of the TF relation.
The amplitude of the scatter depends on the range of $\lambda$.
The range $\lambda=0.04-0.10$, provided by the cosmological tidal field,
results in the appropriate scatter.
\label{fig:tf_ma}}
}
\end{figure}

{\it The slope}:
As seen in Figure \ref{fig:tf_comp}, our simulations well reproduce the slope
and scatter of the TF relation. The shifted zero point may not affect the
mutual comparison among the simulated galaxies. In the left panel of
Figure \ref{fig:tf_ma},
the simulated galaxies with different masses are represented by different
symbols on the TF plot. The slope of the solid line is derived from observed
data (Giovanelli et al. 1997a).
Figure \ref{fig:tf_ma} clearly shows that larger total mass of galaxies
result in brighter luminosity and larger line-width. The resultant slope
of model galaxies on the TF plot is quite similar to the observations.
The TF correlation originates in the differences of total masses of galaxies.

{\it The scatter}:
Figure \ref{fig:tf_ma} (left panel) shows that $I$-band luminosity are
different among the galaxies with the same mass. The reason of the non-constant
$M/L$ is found in the right panel of Figure \ref{fig:tf_ma}. The right panel
is the same plot as the left panel, but different symbols are used for
different initial spin parameters $\lambda$. It shows that galaxies with
different $\lambda$ distribute nearly perpendicular to the TF correlation,
resulting in the scatter of the TF relation. The amplitude of the scatter
depends on the range of $\lambda$. We took $\lambda=0.04-0.10$,
which is a range if angular momenta are caused by the cosmological tidal field
(Barnes \& Efstathiou 1987).
The role of spin parameters on the scatter can be explained in two ways:
(i) Low (high) $\lambda$ leads to a centrifugally concentrated disk, which
provides large (small) line-width $W_{20}$.
(ii) Low (high) $\lambda$ results in smaller (larger) disk, and high (low)
surface density of the gas disk, which controls the time-scale of star
formation $\tau_{\star}$. For low $\lambda$, $\tau_{\star}$ is relatively
short due to the high surface density, and most of stars are formed
at the earlier phase of galaxy formation. Therefore, the mean age of
the stellar component in low $\lambda$ galaxies is older than that
in high $\lambda$ galaxies at $z=0$. Low $\lambda$ makes fainter galaxies
than high $\lambda$ through the age of their stellar component.
This effect also makes the scatter of the TF relation.

\section{SUMMARY AND DISCUSSION}
We have investigated the origin of the TF relation by simulating formation
of 14 galaxies with the $N$-body/SPH direct calculations, which includes
the cooling, star formation and stellar feedback. Internal structures
of spiral galaxies, such as the exponential disk profile and flat rotation,
are well reproduced as the results of the simulations, while these properties
were assumed {\it a priori} in past semi-analytic approaches
(Mo, Mao \& White 1998; Heavens \& Jimenez 1999).
We {\it observed} the total magnitude and gas line-widths of
the simulated "spiral galaxies", and found that they also reproduce
the slope and scatter of the TF relation, except for the zero-point.
We have simulated galaxy formation from many different initial conditions,
and found that the slope of the TF relation is produced by the difference
of total galactic masses, and the scatter is produced by the difference of
initial
spin parameters.

We should comment upon our model of star formation and feedback.
The following issues about star formation and feedback remain for
future investigations with higher resolution simulations.
Luminosity in red passbands is determined firstly by the stellar amount,
and secondly by their age. Simulated galaxies have the stellar amount
similar to the observations, typically $\sim 80-90\%$ of the total baryonic
mass at $z=0$. Their luminosity hence depends on stellar age distribution,
i.e., star formation history. Our star formation model is based
on the Schmidt's law ($\dot{\rho_{\star}} \propto \rho_{\rm gas}^{1.5}$),
which is an optimal but robust model if the mean rate of star formation can
be determined by a mean local density of the gas.
This averaging treatment of star formation would be most suitable for 
the current resolution $\sim 1 \kpc$.
In the Milky Way Galaxy, star formation is observed at molecular clouds,
and further discussion about different kinds of star formation model would
requires to resolve at least the scale of molecular clouds
$\lesssim 100 {\rm pc}$, 10 times smaller than our spatial resolution.
In our simulations, the star formation timescale is mostly set to
$20 \tau_{\rm dyn}$.
Since the local dynamical timescale $\tau_{\rm dyn}$ is much shorter than
galaxy age, the gas is transformed to stars efficiently in high redshift.
Taking the model which gives longer timescales of star formation would
provide brighter galaxies because galaxies which consist of younger stars
are brighter.

Stellar feedback would control the self-regulation mechanisms in
spiral galaxies as discussed by Silk (1997). Our adopted model behaved
like a minimal feedback, because feedback energy released as thermal energy
which could be radiated away in a short timescale.
If the energy is released as kinetic form as discussed in Navarro \& White
(1993), the feedback would become more efficient and could change the history
of star formation. It however is not clear that feedback through kinetic energy
can be occurred in the scale of $\sim 1 {\rm kpc}$, i.e., our spatial
resolution. Since supernovae remnants expand to about $10-100 {\rm pc}$ in the
inter stellar medium, $10-100$ times higher spatial resolution than the present
one should be required for the detailed modeling of the local stellar feedback.
These issues remain for future investigations.

As discussed in Section \ref{sec:TF}, the discrepancy of the $1.5 \magni$
fainter zero point in $I$-band would be alleviated in the cosmological model
of $h>0.5$ and $\Omega_0<1$. We should note, however, that this discrepancy
cannot be solved in the $h=0.5$ and $\Omega_0=1$ cosmology with different
models of star formation and feedback. The $I$-band luminosity declines
by only $1.6 \magni$ from $z=3$, the time of the peak luminosity,
to the present. In order to shift the zero point $1.5 \magni$ brighter
by changing the star formation history, almost all the stars must be formed
at very low redshift $z<0.1$ (recent 2 Gyr) in all the spiral galaxies
in the local universe.

Our initial conditions consider collapse of isolated spheres.
Under these idealized initial conditions, the infall of sub-galactic clumps
at $z \lesssim 2$, which is expected in the $\Omega=1$ CDM universe, cannot
occur, though a continuous gas supply would enhance the star formation rate
at a low redshift. If we consider later gas infalls, the resultant galaxies
would have brighter luminosity. Note however that as discussed
in Section \ref{sec:TF}, our derived TF relation was consistent
with the previous simulations (Steinmetz \& Navarro 1999),
which included the environmental effects, e.g., tidal field and infall/outflow
of mass. In our isolated conditions, the simulations do not suffer from
an extreme transfer of angular momentum from baryon to dark matter
(Navarro, Frenk \& White 1995; Navarro \& Steinmetz 1997) because no merging
occurs at a low redshift,

\acknowledgments
Numerical computations were carried out on the GRAPE-3 system at
the Astronomical Data Analysis Center of the National Astronomical
Observatory, Japan. J.K. thank the Hayakawa Fund for the financial
support to participate the 15th IAP meeting. 
We would like to thank Dr. N. Arimoto for providing us with their
tables of the stellar population synthesis. We thank M. Honma for
reading the manuscript and giving helpful comments.

\clearpage

\clearpage

\clearpage
{\epsscale{0.45}
\plotone{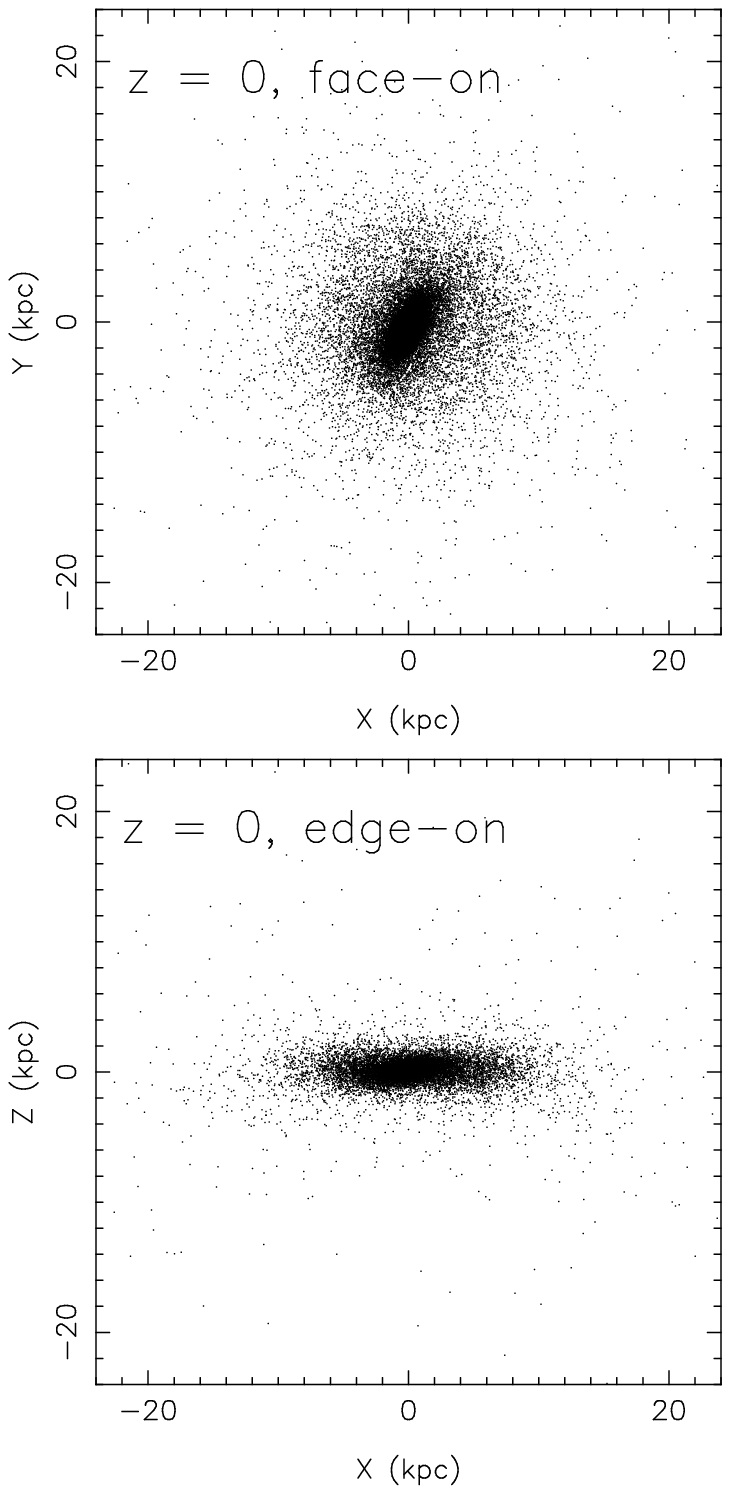}
\figcaption{Snapshots at $z=0$ of star particles in the case for
$M=4 \times 10^{11} \Msun$ and $\lambda =0.06$.
{\it Upper panel:} a face-on projection.
{\it Lower panel:} a edge-on projection.
Note that this plot does not actually show the mass density because
each star particle does not necessary have the same mass.
\label{fig:snap_z=0}}
}

%
\clearpage
\clearpage
\clearpage
\clearpage
\begin{table}
\begin{center}
\begin{tabular}{ccccc}
\hline
\multicolumn{1}{c}{ } & \multicolumn{1}{c}{$E_{\rm rls}$} & \multicolumn{1}{c}{$M_{\rm rls}$} & \multicolumn{1}{c}{$Mz_{\rm rls}$}      & \multicolumn{1}{c}{Period} \\
 & \multicolumn{1}{c}{\small{(erg/star)}} & \multicolumn{1}{c}{\small{($\Msun$/star)}} & \multicolumn{1}{c}{\small{($\Msun$/star)}} & \multicolumn{1}{c}{\small{($10^8$ yr)}}  \\
\hline \hline
SW    & $10^{49}$  &  ---  & --- & $0-0.4$ \\
SNIa  & $10^{51}$  &  1.4  & 1.4 & $0-0.4$ \\
SNII  & $10^{51}$  & 15.3  & 2.5 & $5-30$ \\
\hline 
\end{tabular}
\caption{A star with $\geq 8 \Msun$ is assumed to release the total energy $E_{\rm rls}$,
mass $M_{\rm rls}$ and metal $Mz_{\rm rls}$ par a star through stellar wind (SW),
type Ia (SNIa) and II (SNII) suparnovae. The values listed above are averaged over stars
distributed according to the Salpeter's IMF, using the yield tables of nucleosynthesis
(Nomoto et al. 1997a \& 1997b). We set the upper (lower) mass cut of the IMF to $60 \Msun$
($0.1 \Msun$). The number of stars with $\geq 8 \Msun$ in a {\it star particle}
can be estimated with the Salpeter's IMF.\label{tab:eject}}
\end{center}
\end{table}
\begin{table}
\begin{center}
\begin{tabular}{cccccc}
\hline
\multicolumn{1}{c}{Run} & \multicolumn{1}{c}{M}
 & \multicolumn{1}{c}{$\lambda$} & \multicolumn{1}{c}{$\delta \rho/\rho$} 
 & \multicolumn{1}{c}{$W_{20}$} & \multicolumn{1}{c}{$M_I$}\\
 & \multicolumn{1}{c}{($\Msun$)} & & & \multicolumn{1}{c}{\small{($\kmps$)}} & \multicolumn{1}{c}{\small{(mag)}} \\
\hline \hline
1  & $8 \times 10^{11}$  & 0.10 & $1.9 \sigma$ & 466 & -22.31 \\
2  &                     & 0.08 & $1.9 \sigma$ & 524 & -22.07 \\
3  &                     & 0.06 & $1.9 \sigma$ & 549 & -21.90 \\
4  & $4 \times 10^{11}$  & 0.08 & $1.7 \sigma$ & 420 & -21.32 \\
5  &                     & 0.06 & $1.7 \sigma$ & 437 & -21.21 \\
6  &                     & 0.04 & $1.7 \sigma$ & 454 & -21.06 \\
7  & $2 \times 10^{11}$  & 0.10 & $1.5 \sigma$ & 299 & -20.78 \\
8  &                     & 0.08 & $1.5 \sigma$ & 331 & -20.66 \\
9  &                     & 0.06 & $1.5 \sigma$ & 342 & -20.47 \\
10 &                     & 0.04 & $1.5 \sigma$ & 349 & -20.34 \\
11 & $1 \times 10^{11}$  & 0.10 & $1.4 \sigma$ & 236 & -20.09 \\
12 &                     & 0.08 & $1.4 \sigma$ & 248 & -19.92 \\
13 &                     & 0.06 & $1.4 \sigma$ & 257 & -19.80 \\
14 &                     & 0.04 & $1.4 \sigma$ & 254 & -19.66 \\
\hline 

\end{tabular}
\caption{A catalog of the simulated galaxies. Our two initial free parameters,
total mass $M$ and spin parameter $\lambda$, initial overdensity
$\delta \rho/\rho$ above the background field and two observable
values of final galaxies, line-width $W_{20}$ and $I$-band total magnitude
$M_I$, are listed.\label{catalog}}
\end{center}
\end{table}

\end{document}